# 'The Curious Case of Aid and Conflict:
## Causal Evidence from Panel Econometrics and Composite Indices'

*(M. Usman Anwar Goraya, 2025)*

## Abstract

This paper investigates the intricate relationship between Official Development Assistance (ODA) and conflict in the top ten aid-receiving countries in Africa from 2009 to 2023. Using multiple approaches, such as an **Ordinary Least Squares** (OLS) regression framework, **Principal Component Analysis** as well as a **Ridge (L2) Regression**, the study attempts to delve into and evaluates whether conflict—*measured through proxies like political stability, governance indicators, and macroeconomic variables*—statistically and significantly influence aid inflows. The findings suggest a complex interplay: as the regression approach changes, so does the variable most highly impacting ODA inflow.

The empirical results, as per this research's estimation, suggest that aid remains strongly associated with poverty, inflation, and fragility, while institutional voice and accountability display a negative association with aid flows in pooled regressions. Fixed-effects estimates, in contrast, indicate positive links between ODA, political stability, and GDP per capita over time, alongside negative correlations between ODA and perceived corruption. Ridge regression confirms the robustness of institutional predictors under multicollinearity and highlights the centrality of governance variables. Overall, donors appear simultaneously responsive to humanitarian need and governance considerations, generating an allocation pattern shaped by what we term the aid–conflict–institutions trilemma: aid is most needed where institutions are weakest and conflict risks are highest, yet those same weaknesses threaten the effectiveness of aid.

The paper contributes to the literature by integrating a rigorous theoretical framework with panel-econometric tools, by synthesizing existing empirical results within a sequence wise identification logic.

# 1. **Introduction**

International Aid forms a large portion of capital transfers and flows to the global south from the developed world. Aid could be defined as any form of assistance, typically as money, resources, or services, provided to individuals, organizations, or countries to support their economic, social, or humanitarian needs (Merriam-Webster, n.d.). Aid can further be categorized into various forms, including financial aid, humanitarian aid, development aid, and military aid. For the purpose of this research, development aid will be focused upon since that has, in theory, some of the most direct economic consequences for a country's populace's benefit at large. This will be the figures of the total *Official Development Assistance* (ODA) received by the countries in question over a certain period, otherwise, known as '*foreign aid*'. Foreign aid remains one of the most prominent instruments through which the international community seeks to reduce poverty, promote growth, stabilize fragile states, and prevent violent conflict. Since independence, many African countries have received substantial volumes of official development assistance (ODA), yet the continent continues to exhibit persistent institutional fragility, recurring episodes of violence, and wide development gaps. This empirical regularity has generated a long-running debate about the precise role of aid in shaping conflict dynamics and institutional outcomes.

Moreover, Oxford dictionary (n.d.) defines conflict as 'prolonged armed struggle'. However, in order to focus the scope of this study, conflict will only be defined, observed and studied through the lens of various development factors which would include statistics from, but not limited to, the Political Stability Index (World Bank Repository, 2023), frequency of armed conflict events (Uppsala Conflict Data Program), and macroeconomic indicators such as (*amongst others*) inflation, corruption perception, democratic ratings, GDP per capita, Trade, Current Account Balance, (as proxies for government capacity and economic markets' stability) and the citizenry's Voice & public Accountability.

The core question motivating this paper is how aid, conflict, and institutions interact in Africa's largest aid-receiving countries. Aid can act as a stabilizing force: it eases fiscal constraints, finances social services and infrastructure, and can strengthen the perceived legitimacy of the state when benefits are broadly shared throughout the country. At the same time, aid can unintentionally fuel conflict by increasing the value of capturing the state, relaxing budget constraints on coercion, entrenching patronage networks that marginalize opposition groups, or by way of sheer fungibility of aid. These opposing channels imply that the sign and magnitude of the aid–conflict relationship is likely to be context-dependent and mediated by institutional quality.

The interaction is further complicated by reverse causality. Conflict itself attracts humanitarian, reconstruction, and stabilization aid. Donors often increase assistance in response to crises, generating a simultaneous relationship between conflict and ODA. Simple cross-sectional correlations therefore provide little guidance on whether aid dampens or exacerbates violence. Modern work emphasizes that the aid–conflict nexus is fundamentally conditional: aid in a strong institutional environment behaves differently from aid in a weak one, and the same volume of ODA can have stabilizing or destabilizing effects depending on governance structures and political incentives.

This paper focuses on the ten largest African ODA recipients between 2009 and 2023, namely, Cote D'Ivoire, Niger, DRC, Nigeria, Egypt, Tanzania, Ethiopia, Uganda, Kenya, and Mozambique. These countries combine large aid volumes, complex conflict histories, and substantial institutional reform efforts. Along the econometric analysis, an Aid Dependence Pressure Index (ADPI) is also proposed to reflect a country in question's fiscal dependence on external resources and the interaction of political and environmental fragility.

The remainder of the paper proceeds as follows. Section 2 reviews the literature on aid effectiveness, conflict dynamics, donor selectivity, and aid dependence, highlighting the existing theoretical discourses. Section 3 develops a conceptual model of the aid–conflict–institutions trilemma and motivates the new composite index. Section 4 describes the data sources, country coverage, and index construction procedure. Section 5 lays out the econometric strategy, including pooled OLS, fixed effects, (PCA), ridge regression, and the logic of dynamic panel models. Section 6 synthesizes and interprets the empirical results from the existing estimations. Section 7 discusses the policy implications of the findings, and Section 8 concludes with suggestions for future research.

## 2. **Literature Review**

The motivation for this research stems from the notion that many lesser developed countries are reliant on foreign development assistance for their economic stability. Economic stability is then argued to flow into peace and calm in the otherwise conflict-ridden zones (Polachek et al 1999). However, there are two sides to this debate in the academic discourse. This section highlights the strands most relevant to our empirical questions and methodological choices, focusing on five themes: aid and growth, aid and conflict, institutions and donor selectivity, aid dependence and taxation, and methodological evolution in aid research.

Raleigh and Hegre (2009) argue that conflicts, in most case, are localized, and hence if the developed world utilizing any economic vessel (its own donor agency, a multilateral development agency, etc.) pours in aid, the conflict settles down. This is also linked to the development activities being labour intensive, steering away the general population from any conflict creating activities, all the while where stability in the social and economic fabric leads to a positive impact of aid on conflict (Iyengar et al 2011). In the same context, a few academics argue that the aid for some developing countries are mere extensions of the donor countries foreign policy and global influence (Apodaca 2017). This leads to the aid once disbursed, not being followed through with and opening it to fungibility (Feyzioglu et al., 1998) marred with governance recipient governmental malpractices. This further exacerbates the conflict instead of alleviating it. Grossman (1992) shows theoretically that foreign aid can raise the incentives for insurrection by increasing the value of state capture. Similarly, Balla and Reinhardt (2008) show that conflict influences foreign aid allocation in donor-specific ways, underscoring that aid flows respond to political and strategic considerations rather than humanitarian need alone. The same kind of an inverse relationship also holds true when the aid and development activities are hindered by conflict groups, further leading a region into more unrest (Crost, Felter, and Johnston, 2014).

The shortlisting of African economies for this econometrics analysis also stems from the initial studies by Van de Walle (2001). Van de Walle (2001) argues that foreign aid and economic reforms repeatedly failed in Africa because the political systems can be argued to be redirecting resources toward elite survival and patronage, producing a 'permanent crisis' equilibrium rather than structural transformation.

A second stream of literature, such as the one by Collier and Hoeffler (2002) examined the relationship between foreign aid and the likelihood of civil war outbreaks. Although their specific empirical research found no direct causal link between economic policy, aid, and conflict; however, a cascading effect was evident. This manifests in how poor economic policies often lead to reliance on primary commodities as the main exports, which in turn increases the risk of conflict. In a subsequent study, Collier, and Hoeffler (2006) observed that foreign aid could contribute to a regional arms race. Their research also empirically disproved the notion that military aid, or official development assistance (ODA) when diverted to military spending due to fungibility, serves as a deterrent against rebellions or civil war (Murshed & Sen, 1995). In fact, they found no evidence that military aid reduces the likelihood of internal conflict. This is why Barbanti Jr. (2006) argues that development aid can support conflict resolution only when it is politically informed and conflict-sensitive, while poorly designed aid risks reinforcing existing tensions.

A third body of work such as one from Tarek and Ahmed (2017) concluded that internal conflict fosters political instability. They argue that conflict attracts aid, but this inflow is not necessarily a remedy for the affected country. Weak rule of law and persistent violence or terrorism create significant fiscal and external imbalances, leading to an economic void that is filled with foreign aid and increasing public debt. This cycle ultimately exacerbates instability rather than resolving it.

However, Tahir (2017) alludes to how there is empirical evidence on the interplay of aid and conflict. In her research, although being specific to a certain South Asian country, and showing one-sided effects of inflowing foreign aid on internal conflict, does lay down serious groundwork for further inquisition into the topic, and to decipher if the relationship has a two-way effect, beyond just the national balance sheet, or not. Similarly, there are other economic factors to be observed in recipient countries as well, apart from their current accounts. In support of this notion, Corden (1984), in their work, demonstrates how large external inflows can distort domestic economies through real exchange rate appreciation, providing a theoretical basis for concerns about aid-induced structural imbalances.

The fourth strand concerns aid dependence and taxation. When governments receive large volumes of aid, they may feel less pressure to mobilize domestic tax revenue and to negotiate the fiscal social contract with citizens. This can weaken accountability mechanisms and entrench elite bargains insulated from electoral pressure (De Ree and Nillesen 2009). Conversely, some donors now explicitly support tax administration reforms and public financial management systems, potentially strengthening institutions. The net impact is an empirical question and likely heterogeneous across contexts. The proposed construction of Aid Dependence Pressure Index (ADPI) is designed to capture the degree to which states rely on aid relative to domestic resources, providing a useful summary measure for this very debate.

This paper is an attempt to try and decipher if the claims that ODA may lead to conflict (*proxy measured by the stability and macroeconomic indicators discussed above*) or the counter claim that conflict attracts ODA (aid) is true. Findings may also have broad insights on how aid, especially in natural resource rich regions like Africa, is a means of larger economies to propagate their economic interest by way of aid diplomacy. Here it is also highlighted that prior research often treats governance quality as static, whereas this research uses a panel approach to account for year-on-year variation.

It is also emphasized that while prior studies have extensively debated the bidirectional nature of ODA and conflict, they often fall short in capturing recent post-2009 dynamics in Africa, a region experiencing both renewed aid flows and persistent subnational conflicts, especially on the front of Foreign Direct Investment (FDI) as countries with excess current accounts like China and others look to invest more in African economies. Furthermore, and as already discussed above, many influential studies rely on military aid data or civil war onset as binary measures (often taken as a simple dummy variable in researches), overlooking the subtleties of economic governance, institutional weakness, and aid fungibility, especially in the top aid recipient countries in Africa. This study improves upon previous work by using a continuous panel dataset across 15 years, focusing on top aid-recipient African states. Lastly this paper also attempts to incorporate newer governance and transparency indicators, and further seeks to disentangling their multicollinearity by means of 'Principal Component Analysis (PCA)' before turning towards panel econometrics, and concluding with an effort to make the coefficients more realistic by way of running a 'Ridge Regression'.

### 3. **Theoretical Framework and Economic Theory**

Evident from the review of the existent literature, the effects of aid and conflict on each other, are two ways. Neither can be judged in solidarity. Their distinctive unison is what makes their study a 'curious one.' This is since studies exist which show that aid (although many of them look at 'military aid' specifically, which is beyond the scope of this research piece) can result in increased conflict due to the additional 'aid' in resources being used for the purposes which are not entirely relevant to the original purpose – the sheer utility of the aid, due to its innate fungibility, being called into question.

Therefore, primarily the economic theory at play and which is to be unpacked is the 'Public Choice Theory.' As aid flows in, the governments which are already ranking quite low on the transparency and democracy front tend to divert this fungible aid towards projects (or even other than the intended projects) to their preferred constituencies as a means to consolidate power.

Further, and in reference to the literature review's citation of Collier and Hoeffler's (2002) findings also suggest, that in case of natural resource-oriented economies, which is predominantly the case of the countries which will be studied in different parts of Africa (primary production economies), the economic concept of the 'Paradox of Plenty' also comes into play. Exports of one commodity keeps an economy in their 'comfort zone' per se, leading to overdependence, and even small shocks in the commodities price and/or international markets trickles down to economic downturn, which leads to civil unrest and conflict – paving way for foreign aid inflowing to attempt and stabilize the country since often the donor country sources precious raw material (metals, ground elements, etc.) from recipient.

Relatedly, another economic theory which can be used to as a prism for this problem could be the 'Economic Dependency Theory' which suggests that foreign aid in already struggling economies tends to make those nations more reliant on foreign aid rather than leading to an internal stability towards self-reliance on the social and economic front. This issue again leads to the same mantra, as to whether aid caused this instability and therefore, conflict, of sorts, or in fact the conflict is what attracted aid in the first place.

From a conceptual framework's perspective, based on the literature available it is founded that aid can stabilize or destabilize. In a stabilizing scenario, governments use assistance to expand social services, rebuild infrastructure, and invest in inclusive economic opportunities. This raises the opportunity cost of rebellion, reduces grievances, and increases the legitimacy of the state in the eyes of citizens. Aid can also finance security-sector reform and peacebuilding initiatives that help end violent conflict and prevent its recurrence. In this case, institutional quality, accountability mechanisms, and capable public administration ensure that aid resources are used effectively and equitably.

In a destabilizing scenario, by contrast, aid expands the pool of rents available to those controlling the state. Where executive authority is weakly constrained and corruption is prevalent, leaders may divert aid to patronage networks, security forces, or personal enrichment. Rival elites observe this and face greater incentives to compete for control of the state, potentially resorting to violence if electoral means are blocked. Under such conditions, aid may inadvertently lengthen or intensify conflict by raising the stakes of political competition. The same volume of aid thus has very different implications depending on institutional context and the strategic environment in which it is deployed.

Therefore, in this paper, rather than conceiving of aid as exogenous finance entering a neutral environment, we explicitly model it as a political resource that affects elite strategies, citizen expectations, and donor behavior.

4. **Data, Variables and defining the Index**

The empirical analysis draws on a panel dataset of ten African countries observed annually from 2009 to 2023. These countries were selected because they rank among the largest recipients of ODA in Africa over this period and because they exhibit rich variation in conflict exposure and institutional trajectories. Focusing on this group ensures that the aid–conflict–institutions nexus is central to their development experience.

The dependent variable is net ODA inflows measured in current U.S. dollars, taken from the World Development Indicators. In robustness discussions we also refer to ODA expressed as a share of GDP, on a per capita basis, and relative to government revenue, reflecting different dimensions of aid intensity. Key independent variables include GDP per capita, consumer price inflation, foreign direct investment inflows, trade openness indicators, current account balances, remittances, and a poverty gap measure, all drawn from the same source. Together these macroeconomic indicators proxy the level of development, external vulnerability, and need for external support.

Institutional quality is captured using the World Governance Indicators, focusing on voice and accountability, political stability and absence of violence, control of corruption, and rule of law. These are widely used composite measures based on expert assessments and surveys. While they are not free from measurement error, they offer a consistent comparative basis across countries and over time. Additional governance-relevant information is drawn from the Corruption Perceptions Index produced by Transparency International, which provides another perspective on perceived integrity in the public sector.

Conflict exposure is measured using information on political stability scores and conflict incidence, consistent with the author's original specification. Rather than relying solely on binary civil war indicators, the analysis employs continuous or ordinal measures of instability, allowing the intensity and persistence of conflict risk to vary over time within countries. This approach is more suitable for capturing the spectrum of fragility experienced across the sample, from low-level political unrest to large-scale violent conflict.

To address multicollinearity among macroeconomic openness variables, we implement principal component analysis (PCA) on exports, imports, foreign direct investment, and the current account balance. The first principal component, which explains most of the joint variance among these indicators, is interpreted as an External Openness Index. This index is used instead of the individual components in some regressions, reducing dimensionality and improving numerical stability without discarding economically meaningful information.

We then turn towards the Aid Dependence Pressure Index (ADPI) proposed in this paper is suggested in order to capture the structural reliance of the public sector on external concessional inflows. The index summarizes three dimensions of aid dependence: (i) aid intensity relative to economic size, measured as ODA/GDP; (ii) the contribution of aid to fiscal capacity, measured as ODA/government revenue; and (iii) the weakness of domestic resource mobilization, proxied by the inverse of tax revenue/GDP. All underlying variables are first standardized to have mean zero and unit variance. The tax variable is sign-reversed so that higher values consistently reflect greater aid dependence. The ADPI will then be computed as the average of the three standardized components, although we also discuss a principal-component formulation in which weights are derived endogenously from the covariance structure of the data. In both cases, higher values denote stronger pressure on fiscal and macroeconomic systems arising from aid dependence.

The purpose of introducing ADPI at this stage is conceptual as well as empirical. The index formalizes an idea widely discussed in the aid dependence literature, but rarely summarized in a single measure suitable for cross-country panel analysis. Given the initial, exploratory nature of this contribution, we treat the ADPI as a first-generation measure that will be refined as additional data, peer feedback, and alternative weighting schemes are incorporated. The present formulation should therefore be viewed as a starting point rather than as a definitive metric. A separate paper (or version of this one, thereof) is currently in preparation is intended to develop the ADPI further, explore alternative normalizations, and formally evaluate its predictive properties.

## 5. **Econometric Strategy**

In order to have a detailed look at the interplay of aid and conflict, it will be reiterated how the respective variables will be measured / proxied to do the relevant empirical analysis.

*Aid* = Official Development Assistance (ODA) received by a country in Year X; and

*Conflict\** = Measured in proxy by looking at various factors [as a proxy for government capacity and economic markets' stability]

*\*The exact list of variables that are being taken as dependent along with their details are as follows:*

GDP_per_cap: GDP per capita (indicator of economic development)
Corruption: Corruption index or score (higher = less corruption)
CPI: Consumer Price Index (inflation measure)
FDI: Foreign Direct Investment
Exports: Export values or % of GDP
Imports: Export values or % of GDP
Macro_rating: Macro-economic risk or rating index
CPIA_Prop_Rts: CPIA - Property Rights and Rule-based Governance
Pol_St: Political Stability
Voice: Voice & Accountability Index
Dem_Score: Democracy Score (e.g., from Polity IV or V-Dem)
Poverty_Index: Poverty gap index

Given the traditional 'reactive' nature of foreign aid, triggered ex-post to an economic debacle and/or conflict, the Dependent variable in this case would be foreign aid. In this way, all factors as enunciated above which represent conflict will become the independent variables.

The starting point is a pooled OLS specification that regresses ODA inflows on a set of macroeconomic, governance, and conflict indicators. This provides a benchmark for understanding the unconditional associations in the data. However, pooled OLS ignores unobserved country characteristics such as geography, state of institutional strength, and so on, that may influence both aid and explanatory variables. To mitigate this, the analysis moves to fixed-effects models that include country-specific intercepts, absorbing all time-invariant differences across countries. Year fixed effects are also included to capture common shocks such as changes in global development norms, shifts in donor priorities, commodity price cycles, and financial crises.

The OLS linear, pooled model would be as follows:

$$\text{Net\_ODA}_{it} \text{ (in Mn USD)} = \beta_0 + \beta_1 \cdot \text{GDP\_per\_cap} + \beta_2 \cdot \text{Corruption} + \beta_3 \cdot \text{CPI} + \beta_4 \cdot \text{FDI} + \beta_5 \cdot \text{Exports} + \beta_6 \cdot \text{Macro\_rating} + \beta_7 \cdot \text{CPIA\_Prop\_Rts} + \beta_8 \cdot \text{Pol\_St} + \beta_9 \cdot \text{Voice}_i + \beta_{10} \cdot \text{Dem\_Score} + \beta_{11} \cdot \text{Poverty\_Index} + \beta_{12} \cdot \text{Remittances} + \beta_{13} \cdot \text{Tax\_Revenue} + \beta_{14} \cdot \text{Voice (Democratic Proxy)} + \epsilon$$

*with*,
$\text{ODA}_t$ being the total foreign aid received in period t,
$\text{ConflictVariable}_{i,t}$ being different measures of conflict as discussed earlier in this section

This basic approach employing the Ordinary Least Squares (OLS) regression model seeks to identify the key economic, governance, and conflict related determinants of net Official Development Assistance (ODA) inflows across countries over time. By setting Net ODA as the dependent variable and including a range of independent variables—such as GDP per capita, corruption levels, CPI, foreign direct investment, export volumes, governance indicators (e.g., voice and accountability, political stability), and socio-economic factors like poverty index, remittances, and income inequality—the model aims to quantify how these factors are statistically associated with the volume of ODA received.

As shown in more detail in the following section, the approach then turns towards a Principal Component Analysis, followed by another Pooled OLS:

$$ODA_{it} \text{ (in Mn USD)} = \beta_0 + \boldsymbol{\beta_1 ExternalOpennessPC1_{it}} + \beta_2 \cdot Corruption + \beta_3 \cdot CPI + \beta_4 \cdot FDI + \beta_5 \cdot Exports + \beta_6 \cdot Macro\_rating + \beta_7 \cdot CPIA\_Prop\_Rts + \beta_8 \cdot Pol\_St + \beta_9 \cdot Voicei + \beta_{10} \cdot Dem\_Score + \beta_{11} \cdot Poverty\_Index + \beta_{12} \cdot Remittances + \beta_{13} \cdot Tax\_Revenue + \beta_{14} \cdot Voice \text{ (Democratic Proxy)} + \epsilon$$

After this, a Country Fixed Effects model is run, with the equation-based depiction as follows:

Country-Fixed Effects (*within*):

$$ODA_{it} \text{ (in Mn USD)} = \alpha_i + \sum_k \beta^k X_{k,it} + u_{it}$$

Lastly, and as can be seen in the results section below, since multicollinearity posed another challenge as seen from the multiple macroeconomic openness variables and governance indicators being highly correlated with each other. High variance inflation factors reported in the original analysis confirm this. To address the issue, the study employs PCA to extract a single External Openness Index from exports, imports, foreign direct investment, and current account indicators. This reduces the dimensionality of the problem and yields a more parsimonious specification. In addition, ridge regression is applied as a penalized estimation technique that shrinks correlated coefficients toward zero, improving the stability of estimates at the cost of introducing some bias. Ridge regression is particularly appropriate when the number of predictors is large relative to the time dimension and when collinearity is known to be severe.

$$\hat{B} \text{ in Mn USD}(ridge) = \arg\min_\beta \sum_{i,t}(ODA_{it} - X_{it}\beta)^2 + \lambda \sum_k \beta^2_k$$

6. **Empirical Results and Interpretation**

The following charts show the overall trend of the ODA flowing in, during the period in consideration i.e. 2009 to 2023 for the 10 countries that are being studied in this paper. Both, the consolidation line graph as well as the country-wise faceted graph show that despite the yearly fluctuations in the ODA amount, overall, the common trend has been upwards over the past 15 years. Some of these countries (especially in North and West Africa) peak around 2011 to 2012 including Egypt and Cote d'Ivoire, which, as per some of the literature available suggests its linkage to the Arab Spring. This also supports the findings we have further discussed in this paper that democratic backsliding leads to more aid, since the aid giving agencies and countries generally support democratic values, and would like to support other countries, such as the aid recipients, to subscribe to the same polity approach.

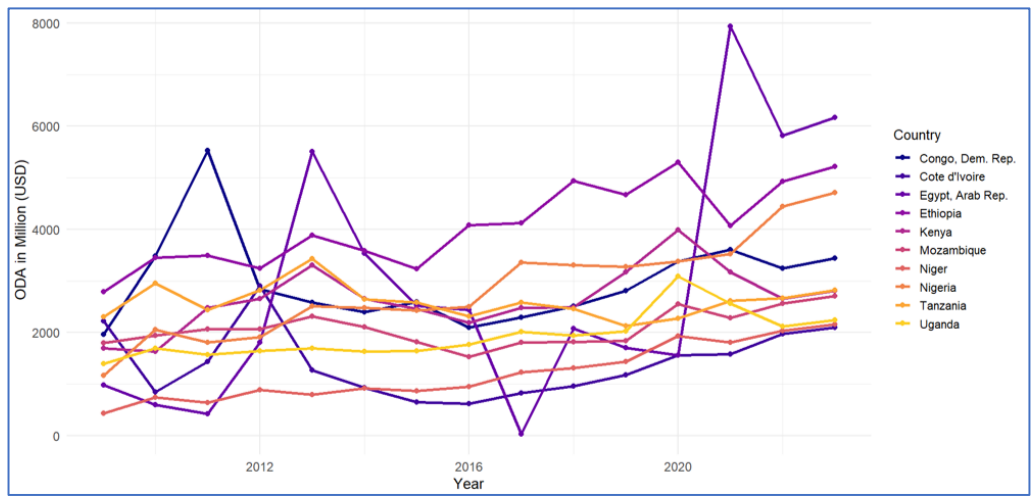

*[Figure 1. Line Graph for the 10 countries – 2009 to 2023]*

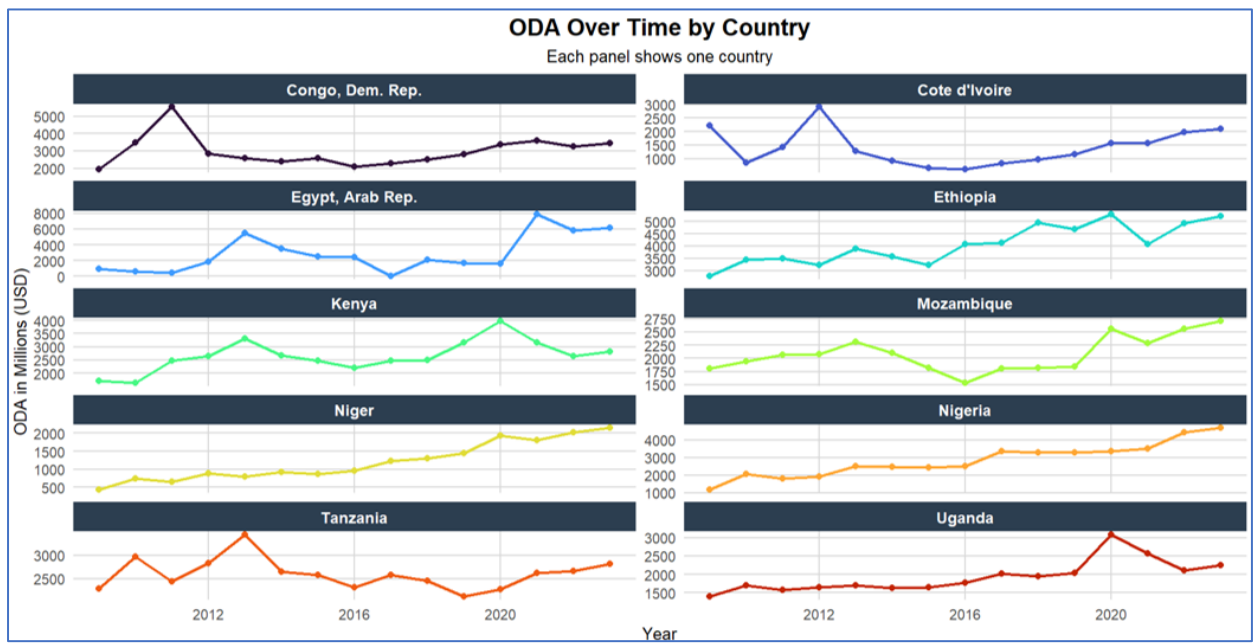

*[Figure 2. Faceted Plot, country-wise – for the years 2009 to 2023]*

Similarly, if we were to look at the raw trends for the different independent variables that are being studied, their distribution can be viewed as follows. This shows the range amongst the 10 countries over the period being studied (2009 to 2023). While variables like the Corruption index remain quite rampant in all 10 countries being studied, others such as the current account (net of export and imports) remain varying in countries. This factor will be discussed in more detail as the paper progresses with its regression analyses and the consequent findings.

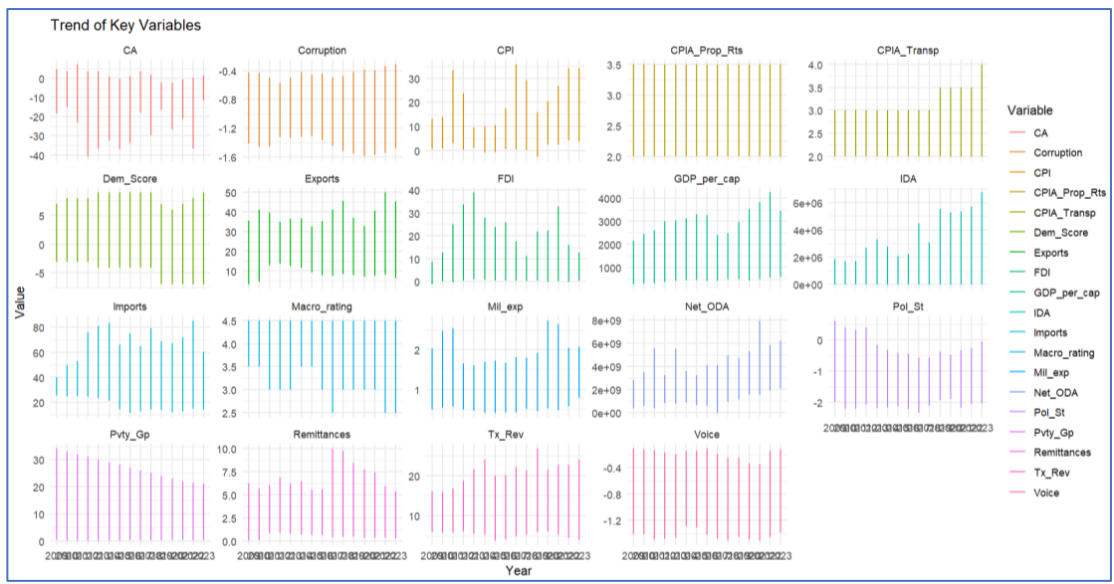

[*Figure 3. Variation of the Independent variables for all 10 countries – net: Years 2009 to 2023*]

For setting a standard, more of a base line, first a pooled, panel wide Ordinary Least Squares (OLS), linear regression was run. This resulted in the following:

| Variable | Estimate | Std. Error | t-Statistic | P-Value | CI Lower | CI Upper |
|---|---|---|---|---|---|---|
| (Intercept) | -2,076.09 | 2,135.95 | -0.97 | 0.33 | -6,301.81 | 2,149.62 |
| 'Current Account' | -0.96 | 29.01 | -0.03 | 0.97 | -58.36 | 56.43 |
| CPI | 41.88 | 14.15 | 2.96 | 0.00 | 13.88 | 69.88 |
| 'Property Rights' | 769.05 | 432.88 | 1.78 | 0.08 | -87.35 | 1,625.45 |
| Transparency | -269.60 | 398.99 | -0.68 | 0.50 | -1,058.95 | 519.75 |
| 'Corruption Index' | -47.72 | 578.60 | -0.08 | 0.93 | -1,192.41 | 1,096.97 |
| Exports | -58.03 | 22.64 | -2.56 | 0.01 | -102.83 | -13.24 |
| FDI | -11.98 | 28.98 | -0.41 | 0.68 | -69.31 | 45.35 |
| GDPperCap | 0.69 | 0.22 | 3.17 | 0.00 | 0.26 | 1.13 |
| Displaced_Persons | 0.00 | 0.00 | 3.90 | 0.00 | 0.00 | 0.00 |
| Imports | 1.07 | 28.06 | 0.04 | 0.97 | -54.45 | 56.59 |
| Macroeconomic_Mgmt | -12.42 | 262.76 | -0.05 | 0.96 | -532.26 | 507.41 |
| 'Military Expenditure' | -221.08 | 265.39 | -0.83 | 0.41 | -746.12 | 303.96 |
| Political_Stability | 377.22 | 282.41 | 1.34 | 0.18 | -181.49 | 935.93 |
| 'Poverty Gap' | 67.25 | 34.44 | 1.95 | 0.05 | -0.90 | 135.39 |
| Remittances | -181.01 | 64.06 | -2.83 | 0.01 | -307.74 | -54.27 |
| 'Tax Revenue' | 118.37 | 46.86 | 2.53 | 0.01 | 25.67 | 211.08 |
| 'Voice and Accountability' | -1,746.96 | 367.17 | -4.76 | 0.00 | -2,473.35 | -1,020.56 |
| 'Democracy Score' | 56.80 | 37.75 | 1.50 | 0.13 | -17.90 | 131.49 |
| Violence_Casualties | 0.00 | 0.01 | 0.03 | 0.98 | -0.01 | 0.01 |

Here, we see that the shaded rows are the statistically significant variables (*at least at the 10% level*). Notably, the Displaced persons, although significant, does not impact any of the ODA flow to any of the countries in consideration. We can also see that Poverty Gap and Tax Revenue (Estimate = 118.37 million USD, *p* = 0.01), are two more significant variables. These are also supplemented by Voice and Accountability (Estimate = -1,746.96 million USD, *p* = 0.04) as well as Remittances being significant as well. Together these variables suggest that the more poor, lesser remittances, lesser open to public opinion and more efficient in tax collection a country is – the more ODA is received by these nations.

However, since most of these variables are macroeconomics based, the interlinkage of the economy may very well have led to correlation between them. We now turn towards having a look at the correlation matrix.

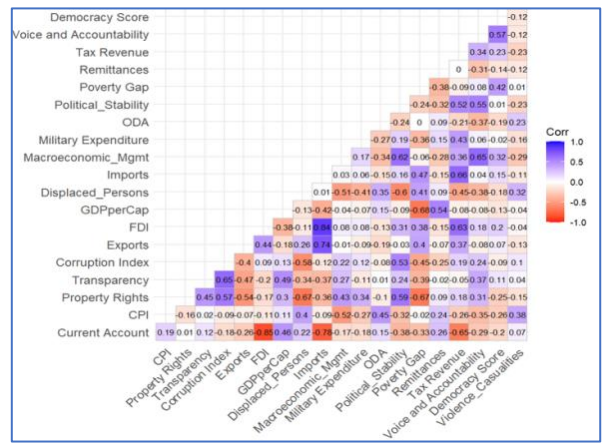

[*Figure 4. Correlation Matrix of the Independent Variables*]

Quite evidently, we see darker shades of blue (high positive correlation) and dark shades of red (high negative correlation). This is certainly an issue that needs to be addressed in order to be able to look at the results of the statistical analysis with more confidence. This is why a cut off for 0.70 (or 70%) correlation in either direction is set, and we turn towards a Principal Component Analysis. This will refine our subsequent regressions by addressing part (if not all) of the collinearity issue. But before that, a Variance Inflation Factor analysis is also done to further see which of these are multilaterally causing multicollinearity.

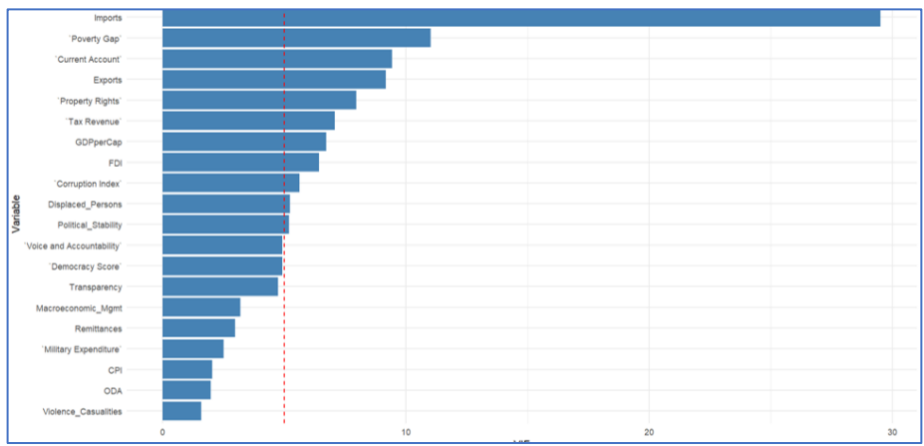

[*Figure 5. Multilateral VIF Chart: Threshold at 5*]

The VIF chart also shows that the variance is being built upon as we are adding more variables in the regression with Imports, Current Account, Exports and FDI, notably being above the threshold level of 5. Since these also have a correlation value (bilateral) of above 0.7 amongst each other, we will analyze them further.

The following Scree Plot shows that 04 of the highest correlated variables, namely Current Account, FDI, Export and Imports (mentioned prior) when run together, gives us 03 possible options to define a Principal Component or an Index.

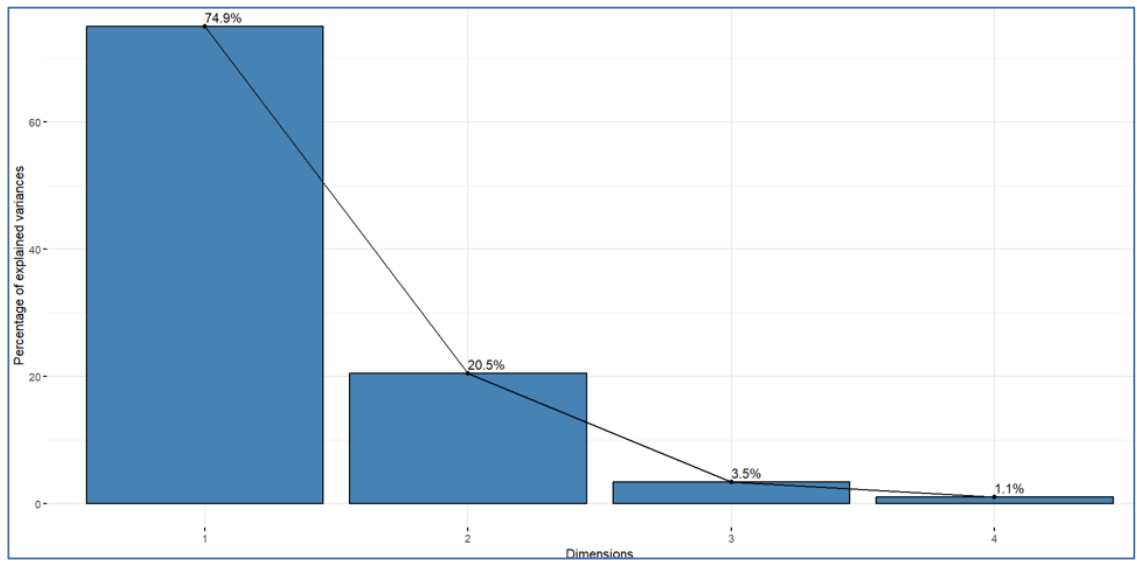

[*Figure 6. Scree Plot for the 04 highest correlated variables*]

Further analysis of the Scree Plot also shows that the 'elbow' that is usually looked for while doing this kind of an analysis can be seen at level 2. Together PCA1 and PCA 2 will then account for **95.4%** (*74.9% and 20.5%*) of the correlation in those 04 variables in consideration. However, as the following details on variable loadings show us, PCA 2 consists largely of Current Account, hence, only PCA1 was selected to attempt and avoid the issue of second tier multi collinearity amongst the two chosen Principal Components.

| Variable Loadings on Principal Components | | |
|---|---|---|
| Variable | PC1 | PC2 |
| Current Account | −0.50 | 0.50 |
| Exports | 0.39 | 0.81 |
| FDI | 0.53 | −0.28 |
| Imports | 0.56 | 0.14 |

[*Figure 7. Variable Loadings for PCA 1 and PCA 2*]

This leads us to further analyze and cross check selection of PCA1, via a load plot, and see how the 04 different variables are impacting the correlation amongst themselves when plotted over the two PCA categories (*labelled as Dimension 1 and Dimension 2 in the plot*).

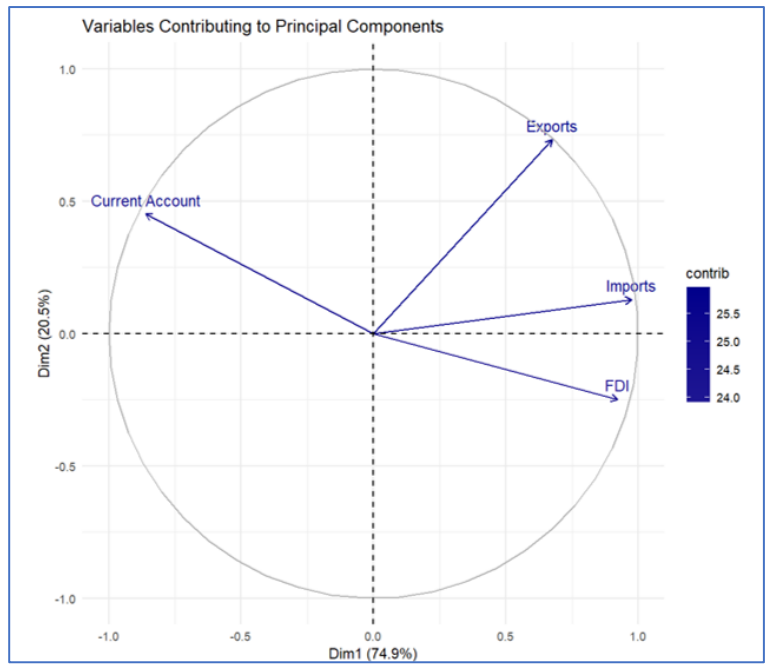

[*Figure 8. Variables feeding into the Principal Components – PCA 1 & PCA 2*]

This plot shows us that while all four are closer to the circumference, showing their strong contribution to the correlation within the index that we are trying to form here (PCA), their directions are such that Export, Imports and FDI move in one direction while the Current Account moves in another. This supplements economic theory as well, since aid dependent countries often import more than they export, and with companies which do end up investing (FDI), often repatriate their profits due to macroeconomic stability issues. All of this together leads to a current account deficit, which is also shown by the load plot above.

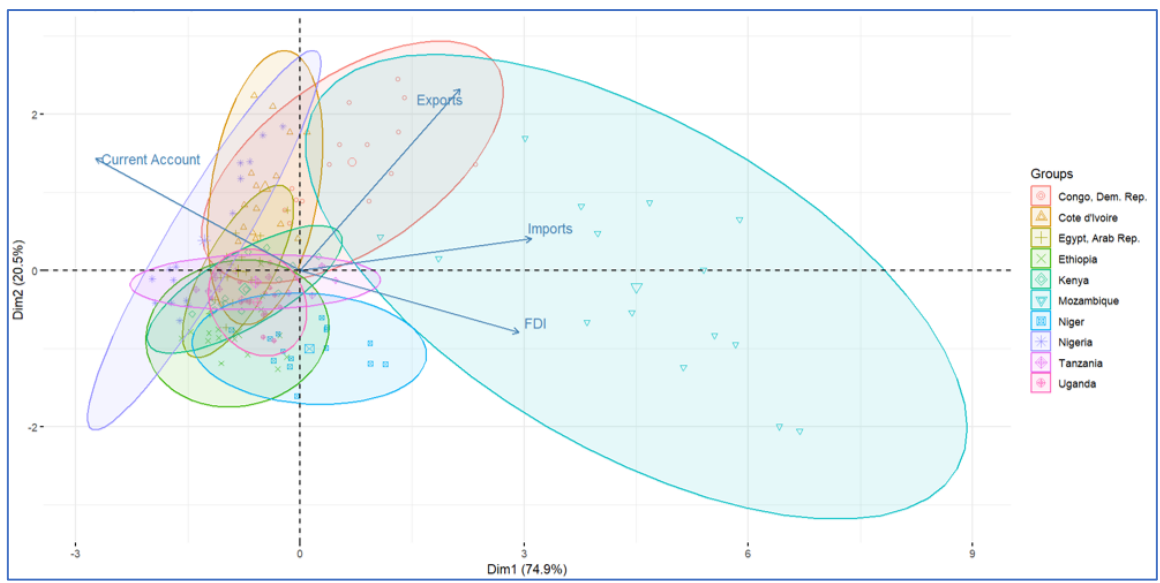

[*Figure 9. Country-wise load plot*: *Principal Component Analysis*]

Subsequently, the country wide load plot also shows the within country interaction of the 04 variables in consideration for having a PCA being indexed. The more *north* and *east* a country is in the graph, the more open to trade it will be – say Mozambique. The inverse is true for countries falling in the *south west* of the plot, which include DRC and Ethiopia. The countries in the middle of the graph will be balance on account of their interaction of FDI, Imports, Exports, and the Current Account.

With all the above information in view, the 04 variables under consideration are then indexed into one Principal Component and it is named 'External Openness'. Doing so will certainly assist further regression-based analysis since it will counter the correlation between the error terms and the coefficients of these variables.

With the PCA Index now formed, we run the linear Pooled OLS one more time to get the following results:

### OLS Regression Results (Significant Results Highlighted)

| Variable | Estimate (Millions) | Std. Error | t-Statistic | P-Value | CI Lower | CI Upper |
| --- | --- | --- | --- | --- | --- | --- |
| (Intercept) | -2,836.15 | 2,101.00 | -1.35 | 0.18 | -6,991.55 | 1,319.26 |
| External_Openness_PC1 | -262.56 | 118.08 | -2.22 | 0.03 | -496.10 | -29.02 |
| GDPperCap | 0.37 | 0.22 | 1.69 | 0.09 | -0.06 | 0.81 |
| `Corruption Index` | -493.07 | 600.00 | -0.82 | 0.41 | -1,679.77 | 693.62 |
| CPI | 46.72 | 14.89 | 3.14 | 0.00 | 17.28 | 76.16 |
| Political_Stability | 268.35 | 287.43 | 0.93 | 0.35 | -300.15 | 836.84 |
| Remittances | -109.51 | 66.48 | -1.65 | 0.10 | -241.00 | 21.97 |
| `Tax Revenue` | 107.55 | 45.48 | 2.36 | 0.02 | 17.60 | 197.51 |
| `Voice and Accountability` | -1,474.39 | 381.78 | -3.86 | 0.00 | -2,229.48 | -719.31 |
| `Democracy Score` | 20.55 | 35.10 | 0.59 | 0.56 | -48.87 | 89.97 |
| Violence_Casualities | 0.01 | 0.01 | 1.49 | 0.14 | -0.00 | 0.02 |
| Macroeconomic_Mgmt | -314.74 | 275.57 | -1.14 | 0.26 | -859.77 | 230.28 |
| `Military Expenditure` | -352.48 | 283.94 | -1.24 | 0.22 | -914.07 | 209.11 |
| Transparency | 251.17 | 417.77 | 0.60 | 0.55 | -575.10 | 1,077.45 |
| `Property Rights` | 660.52 | 362.16 | 1.82 | 0.07 | -55.78 | 1,376.82 |
| `Poverty Gap` | 77.98 | 36.66 | 2.13 | 0.04 | 5.48 | 150.47 |

These results now show us that CPI has a strong positive effect ($\beta = 46.72$ million USD, $p < 0.01$) on ODA, indicating that inflationary trends are associated with higher inflows. Moreover, Voice and Accountability show a large negative impact ($\beta = -1,474.39$ million USD, $p < 0.01$), suggesting that weaker institutional voice reduces the aid flowing into a country by around $1.5 Billion per dropped index point in Voice and Accountability. Quite interestingly, the index, that was constructed as a result of the Principal Component Analysis, External Openness (PC1), is also negatively associated ($\beta = -262.56$ million USD, $p = 0.03$) with aid. This implies that more trade independent, and open economies tend to receive fewer aid money. Moving to Tax Revenue, we see a healthier tax collection system contributes positively ($\beta = 107.55$ million USD, $p = 0.02$) to aid inflow, emphasizing the importance of domestic fiscal strength. And lastly in this case, Poverty Gap is positively significant ($\beta = 77.98$ million USD, $p = 0.04$), indicating areas with wider poverty may attract higher flows.

However, countries have unobserved, time-invariant unique characteristics which may not change over time, we now turn to deciphering the regression with these effects in mind. This will be done by employing the Country Fixed Effects approach. This will not only improve our causal interpretation between the impact of ODA and the proxied variables representing conflict, but will also discount the omitted variable bias that may lurk in our analysis, effectively looking to treat endogeneity.

When the said regression was run with Country-Fixed Effects, we see the following results:

| Variable | Estimate (Millions) | Std. Error | t-Statistic | P-Value | CI Lower | CI Upper |
|---|---|---|---|---|---|---|
| External_Openness_PC1 | -28.91 | 129.19 | -0.22 | 0.82 | -282.12 | 224.30 |
| GDPperCap | 1.93 | 0.35 | 5.47 | 0.00 | 1.24 | 2.62 |
| Corruption.Index | -1,567.03 | 756.47 | -2.07 | 0.04 | -3,049.69 | -84.37 |
| CPI | 23.70 | 14.08 | 1.68 | 0.09 | -3.90 | 51.31 |
| Political_Stability | 872.12 | 352.24 | 2.48 | 0.01 | 181.74 | 1,562.50 |
| Remittances | -9.75 | 85.17 | -0.11 | 0.91 | -176.69 | 157.18 |
| Tax.Revenue | 58.74 | 62.56 | 0.94 | 0.35 | -63.88 | 181.36 |
| Voice.and.Accountability | -399.55 | 537.99 | -0.74 | 0.46 | -1,453.99 | 654.90 |
| Democracy.Score | 60.76 | 35.98 | 1.69 | 0.09 | -9.76 | 131.29 |
| Violence_Casualities | 0.00 | 0.01 | 0.24 | 0.81 | -0.01 | 0.01 |
| Macroeconomic_Mgmt | -723.93 | 285.41 | -2.54 | 0.01 | -1,283.33 | -164.53 |
| Military.Expenditure | 329.49 | 269.73 | 1.22 | 0.22 | -199.17 | 858.15 |
| Transparency | -258.85 | 469.39 | -0.55 | 0.58 | -1,178.84 | 661.15 |
| Property.Rights | 355.96 | 427.19 | 0.83 | 0.41 | -481.31 | 1,193.23 |
| Poverty.Gap | -27.31 | 57.48 | -0.48 | 0.64 | -139.97 | 85.36 |

Country FE Regression Results (Significant Results Highlighted)

Once the country level fixed effects regression was executed, we see in the results that GDP per capita emerges as a highly significant and positive predictor ($\beta = 1.93$ million USD, $p < 0.01$) of ODA inflow, indicating that rising national income levels are strongly associated with increased values of receiving ODA from donor countries and organizations. Similarly, political stability shows a substantial positive impact ($\beta = 872.12$ million USD, $p = 0.01$), reinforcing the importance the donors lay on stable and good governance of recipient countries. However, and conversely, corruption demonstrates a large and statistically significant negative effect ($\beta = -1,567.03$ million USD, $p = 0.04$), suggesting that worsening perceptions of corruption within a country over time are associated with declines in donors providing ODA. Moreover, macroeconomic management also shows a significant negative association ($\beta = -723.93$ million USD, $p = 0.01$), which may reflect transitional costs of structural reforms or fiscal tightening. However, this requires further investigation in ensuing, more detailed research, since this may be due to endogeneity in the data or an outright omitted variable bias. Then, and at a glimpse, while CPI and democracy score approach significance ($p = 0.09$, statistically significant at a 10% level), other variables such as tax revenue, remittances, and voice and accountability lose their explanatory power in the fixed effects model (in comparison with the pre and post principal component analysis simple OLS models discussed prior), implying that their influence is largely driven by cross-country variation rather than time-series trends within countries.

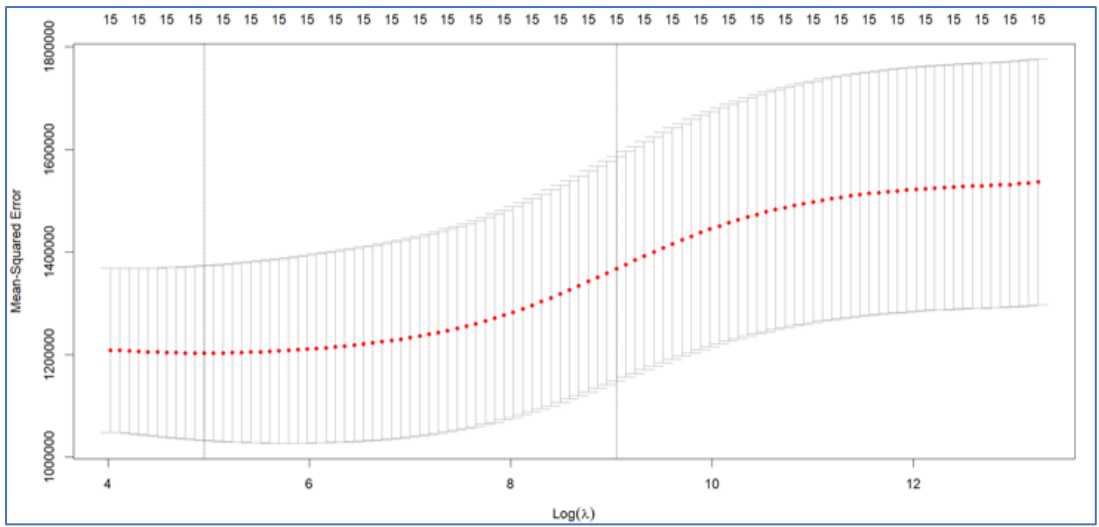

[*Figure 10. Ridge Regression:* **λ** *selected as a conservative log 5*]

Due to persistent multicollinearity, the research would have a limited empirical effectiveness in capturing the true effects of conflict on ODA, if one were to limit to the OLS Pooled Regression and the Country Fixed Effects Regression models. Therefore, we now turn towards a Ridge Regression. This approach is employed in order to reduce inflated standard errors (reference to the VIF chart shown in this section prior), as well as adding to the precision of the prediction coefficients. Ridge Regression will effectively shrink the coefficients, discounting them to a value closer to zero. The coefficients which persist despite the ridge effect, will show better predictive power of the impact of conflict on ODA.

The diagram above shows the cross validation of the mean squared errors across varying levels of log of lambda (which is the regularization factor, otherwise also called the penalty factor). It is a tradeoff between making the mean squared errors (MSE) higher against choosing the right penalty factor. In other words, a tradeoff between bias and variance. The minimal lambda is chosen to be able to incorporate Ridge Regression approach but in view with a modestly aggressive approach; which will avoid over-fitting of the model while steering clear of under-fitting which can occur if the lambda chosen is too high.

Once the penalty factor is applied, the Ridge Regression gives the following output which show that the coefficients in Millions of US Dollars, have shrunk substantially. Since a Ridge Regression does not have p-values, we can interpret the coefficients on their magnitude and direction only. These results reflect that Voice and Accountability retain a large negative coefficient (*–950.69 million USD*), suggesting that diminished institutional freedoms are strongly associated with lower ODA, even after penalization. Similarly, Military Expenditure (*–389.42 million USD*) and Corruption Index (*–300.71 million USD*) show substantial negative weights, highlighting the detrimental role of governance and security inefficiencies on ODA inflows. On the positive side, Property Rights (*318.95 million USD*) and Transparency (*163.97 million USD*) carry strong positive associations, underscoring the importance of institutional clarity and rule of law, which are weighted by donor countries or multilateral agencies and are essentially, as shown here, enabling factors for ODA.

GDP per Capita, however, is nearly neutral (*0.18 million USD*), indicating that once multicollinearity is accounted for, income level adds little incremental predictive power. This also means that in future research this could be used as a good control variable while doing the cross time Fixed Effects. Other moderate contributors include Political Stability (increase of ODA by *98.41 million USD* per index point increase) and Tax Revenue (increase of ODA by *44.66 million USD* per index point increase), which reinforce their directional relationship with ODA inflow, as already seen in the Pooled OLS as well as the country Fixed Effects models above.

| Variable | Estimate |
|---|---|
| (Intercept) | 52.47 |
| External_Openness_PC1 | -116.33 |
| GDPperCap | 0.18 |
| Corruption.Index | -300.71 |
| CPI | 43.29 |
| Political_Stability | 98.41 |
| Remittances | -63.45 |
| Tax.Revenue | 44.66 |
| Voice.and.Accountability | -950.69 |
| Democracy.Score | 12.58 |
| Violence_Casualities | 0.01 |
| Macroeconomic_Mgmt | -197.20 |
| Military.Expenditure | -389.42 |
| Transparency | 163.97 |
| Property.Rights | 318.95 |
| Poverty.Gap | 27.15 |

### 7. **Policy Implications**

The empirical patterns discussed above have several implications for the design and management of aid in Africa's most aid-dependent countries. For donors, the coexistence of need-based and governance-based allocation criteria suggests that programming strategies must explicitly confront the aid–conflict–institutions trilemma rather than treating need and institutional quality as independent objectives. In practice, this means that when donors increase aid to fragile and institutionally weak environments, they should simultaneously invest in transparency, oversight, and local accountability mechanisms that reduce the risk of capture.

Strengthening country systems and local institutions is central to this agenda. Budget support and large-scale government-implemented projects may be appropriate where public financial management systems are robust and oversight bodies are effective, but in more fragile contexts donors may need to rely on hybrid modalities that combine government, nongovernmental, and community-driven structures. Regardless of modality, transparent reporting, third-party monitoring, and inclusive participation, especially by and of marginalized groups (Boone, 1996), can help mitigate the distortions created when large aid flows interact with weak institutions.

Moreover, recipient governments also face important choices. High aid dependence, can create vulnerabilities if donors suddenly shift priorities or reduce funding. To manage this risk, governments should prioritize domestic resource mobilization, strengthen tax administration, and build buffers that reduce the volatility of public spending. Investing in rule of law, anti-corruption efforts, and property rights can enhance both the effectiveness of aid and the resilience of the domestic economy. When governments demonstrate credible reform efforts, they may attract more predictable and flexible aid, creating a virtuous circle between governance improvements and external support.

## 8. Conclusion and Way Forward

This paper has revisited the complex relationship between aid, conflict, and institutions in ten of Africa's largest aid-receiving countries over the period 2009–2023. Building on an existing empirical analysis, it was attempted to showcase the relation that acute conflict as well as chronic macroeconomic instability in the countries in question.

The findings underscore three broad conclusions. First, aid allocation to the countries studied is strongly associated with indicators of poverty, macroeconomic stress, and fragility, confirming that donors continue to respond to need. Second, governance indicators, particularly political stability and corruption control, matter for aid flows once unobserved country characteristics are accounted for, indicating that donor selectivity based on institutional quality is not merely rhetorical. Third, the interaction between aid dependence, institutional incentives, and conflict requires careful attention. In more step-by-step detail it can be seen that, with the Pooled OLS model initially showing that variables such as Displaced Persons, Tax Revenue and Voice & Accountability have significant impact on ODA, a high multicollinearity (confirmed by high VIF scores and bivariate correlation evident in the correlation matrix) led to a Principals Component Approach. When an 'External Openness' index was formed consisting of highly correlated variables of FDI, Exports, Imports and Current Account, we ran a rather more robust Pooled OLS regression. This also confirmed that the Trade Openness and CPI both have implications on ODA inflows, negatively and positively, respectively. We then turned towards a Fixed Effects (within country) model, which reflected that Corruption Index, GDP per capita, Political Stability and Macroeconomic Management were the key variables in determining the amount of foreign ODA inflows for these countries in consideration. While GDP per capita was positively related, the other three were negatively impacting ODA inflows. Lastly due to persistent multicollinearity, a Ridge Regression was run with a modest approach to the penalty factors ($\lambda = \log 5$) in order to decrease the standard errors and make the coefficient estimates more robust. This also confirmed that institutional quality, macroeconomic discipline as well as good governance, together are the variables which if not in sync with each other will cause conflict, and ultimately be impacting the ODA inflows.

The analysis also highlights important limitations. Without experimental or quasi-experimental variation, the econometric strategy cannot fully disentangle causality from correlation. Governance indicators are measured with noise, and conflict dynamics are influenced by many factors outside the model. Despite this, the study offers a more structured account of the aid–conflict–institutions nexus than is often found in similar cross-country analyses.

Future research can build on this foundation in several directions. For example, it would be even more apt to supplement the Country level Fixed Effects with Time Fixed Effects as well as a Random Effects model. A Haussman test may also be conducted in future researches to confirm which one of all three of these effects-based models is more apt and fitting to the conflict proxies and their cumulative as well as independent impact on ODA inflows. A Granger Causality test or a Dynamic Panel Model with more years in consideration, given the large number of dependent variables may also be run which will lead to unpacking the two-way impact of ODA and conflict (*proxies, in this case*) have on each other. Further, the index proposed will also be a way of ranking aid dependencies and the effects it has on the recipient country's institutional and economic strength.

In sum, the paper argues that the central challenge in contemporary aid policy is not simply to increase or decrease aid volumes, but to align aid with institutional reforms and conflict-sensitive strategies that acknowledge the political realities of recipient countries.